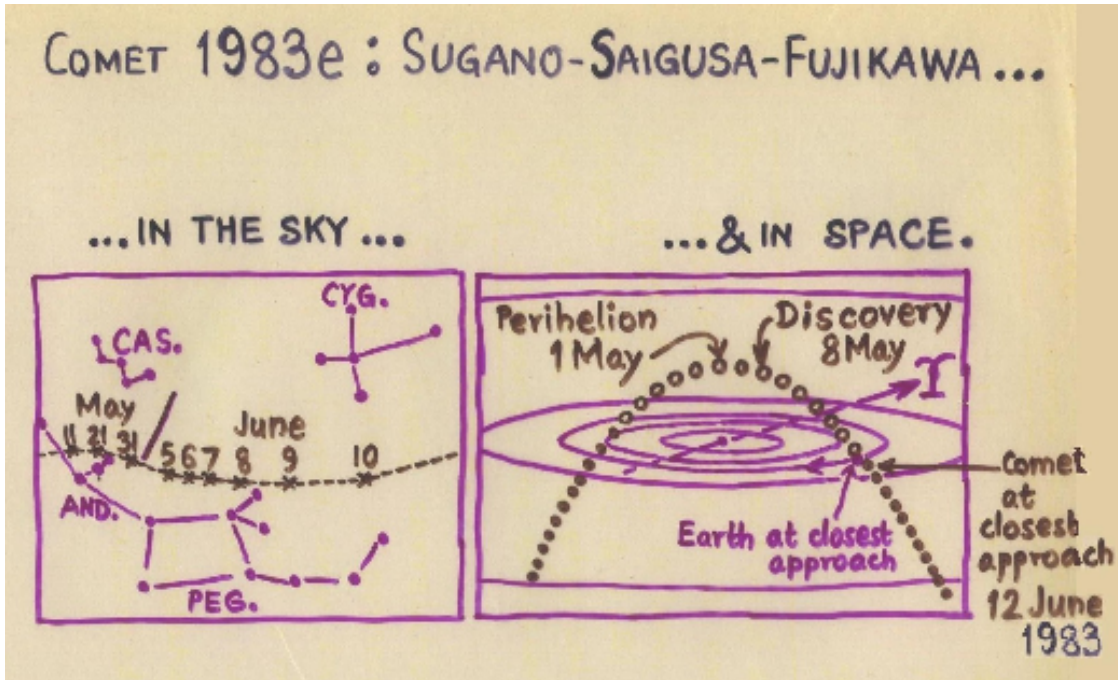

# Occultation of the radio source 2019+098 (3C411) by comet 1983e (Sugano-Saigusa-Fujikawa)

[Brief title: Occultation of 2019+098 by comet 1983e OR (briefer) 1983e cometary occultation]

*by*

**S. Ananthakrishnan & Dilip G. Banhatti**

Radio Astronomy Centre (TIFR), Ooty and Bangalore

[1983]

[This cometary occultation observation from June 1983 remained to be formally reported due to other preoccupations of the authors. It was presented in seminars to colleagues at Ooty, Bangalore and elsewhere. We now write it up as we have been asked about it by various colleagues at various times, and feel we owe it to them to put it firmly on record.]

## Abstract

We planned and observed with Ooty Radio Telescope the occultation with Comet 1983e Sugano-Saigusa-Fujikawa of the extragalactic radio source 2019+098 ≡ 3C411. The results are presented formally for the first time, along with a brief account of other cometary occultations and general background of planning, execution and interpretation of such observations which will be useful for other future observers. The occultation occurred at 07:52 IST on 12[th] June 1983. It amounted to 25% peak to peak fluctuation in the flux density of the radio source. The rough predicted occultation time was 07:24 IST. We interpret the results after refining the occultation time to allow for various effects.

## 1. Introduction

Since 1977, we have kept an eye open for comets coming close enough to the earth to substantially occult background radio sources. The comet IRAS-Araki-Alcock (1983d) was bright enough, but was unfortunately not in the sky over Ooty during its perihelion passage. The opportunity to try to observe another cometary occultation after that by Kohoutek (1973f) in 1974 (Ananthakrishnan et al 1975) came with the discovery of the comet Sugano-Saigusa-Fujikawa (1983e) in May 1983 (IAUC3803).

Successful observation of a cometary occultation is difficult because the ephemeris is not known with sufficient accuracy well in advance. We plotted the path of 1983e from the ephemeris given in IAUC3812, along with radio sources stronger than 2 Jy at 408 MHz from the Molonglo (Large et al 1981), Parkes (CSIRO 1969) and 4C (Pilkington & Scott 1965) catalogues. Out of eight sources observable in the Ooty sky at their nearest approach to the comet (see Appendix 1), two (2019+098 ≡ 3C411, about 10 Jy, occultation expected on 12th June, and 1909-058, about 3 Jy, occultation expected on 13th June) seemed promising and we decided to track these for about two hours straddling the predicted occultation time. Occultation was clearly detected near the (by then updated) predicted time for 2019+098 on 12 June 1983.

We present a description of the observation procedure and precautions taken in section 2 below, the result of the observation in section 3 and derive the solar wind velocity required for the occultation to occur at the observed time, based on the precise ephemeris of 1983e, in section 4. Section 5 discusses possible interpretations of the occultation profile.

## 2. Observation procedure and precautions

Before planning observation, one must make sure that the comet has a plasma tail (type I) & not dust tail (type II) – comet 1983e has a plasma tail (type I). Magnitude at the time of observation should be brighter than ≈ 5$^m$. For 1983e IAUC3820 gives the estimates 6.9 mag on 8th June to 7.3 mag on 17th, brightest being 5.3 mag around 13th June. So it is just bright enough. Could have been better. Comet-Sun distance should be within 1 AU. This condition is met around 13th June. The source position on the sky should be on the comet-Sun line, away from Sun & as close to comet as possible, allowing a small distance toward Sun also – to take care of erroneous ephemeris. Sources were selected with this criterion in mind – see Appendix 1.

### 2.1 The observing system

The observation was done with the Ooty radio telescope (ORT) mainly in its correlated mode. The telescope is an equatorially mounted cylindrical semi-paraboloid having reflecting surface made of stretched stainless steel wires of N-S extent 530 m strung on parabolic trusses of E-W extent 30 m. The focal line has a phased array of dipoles illuminated by the semi-paraboloid. By phasing the dipoles in a N-S gradient, 12 N-S beams of extent ½ degree E-W and 3 arcminutes N-S are formed, adjacent beams spaced roughly a beamwidth apart in N-S. This is called the correlated mode. It is also possible to form 12 so-called total power beams of about double the N-S extent by combining differently the 24 N-S telescope modules. The narrower more sensitive correlated beams drift more than the wider less sensitive total power beams, which are thus stabler. ORT operates at 326.5 MHz (or equivalently λ91.8 cm) with bandwidth 4 MHz.

In our observation, the seventh beam tracked the source position, and four adjacent beams were also used; sixth and eighth for nulls and fourth and ninth for about 40% negative signal. Beam one was also used as off-source monitor and beam 11 in the total power mode to monitor any fast interference. Signals obtained from the correlated beams after integration with a time-constant of 0.5 sec and the total power beam after an integration of 22 msec were displayed on charts with a pen-recorder. In addition, signals from beam seven (main signal), beams five and nine (40% negative signal) and beam one (off-source monitor) were acquired on a magnetic tape after integration with a time-constant 50 msec and analog-to-digital conversion.

## 2.2 The observation sequence

3C435 (7.8 Jy) was used for flux density calibration at the beginning and end. 2019+098 ($\equiv$ 3C411) was tracked continuously with correlated beam 7 on the source position and the adjacent correlated beams 5, 6 and 8, 9 were recorded for the beamshape, while the distant beams 1 in correlated mode for off source level and 11 in total power mode were also recorded to monitor any fast, possibly local, interference. All the beams had stable signal without drifts, both for the calibrator 3C435 and the target source 2019+098. Ionospheric scintillation was monitored by observing 3C435 and 2145+15 (9.0 Jy) before the occultation observation and 3C435 afterwards. No ionospheric scintillation was present, as shown by the stability of the signal. (Ionospheric scintillation has a timescale of variation of the order of minutes, when present.)

The radio source 1909-058, three times weaker than 2019+098, was also observed on 13[th] June, using 1949+023 ($\equiv$ 3C403) for calibration and monitoring ionospheric scintillation before and after the expected occultation observation. No occultation was detected for 1909-058.

------------------------------------------------------------

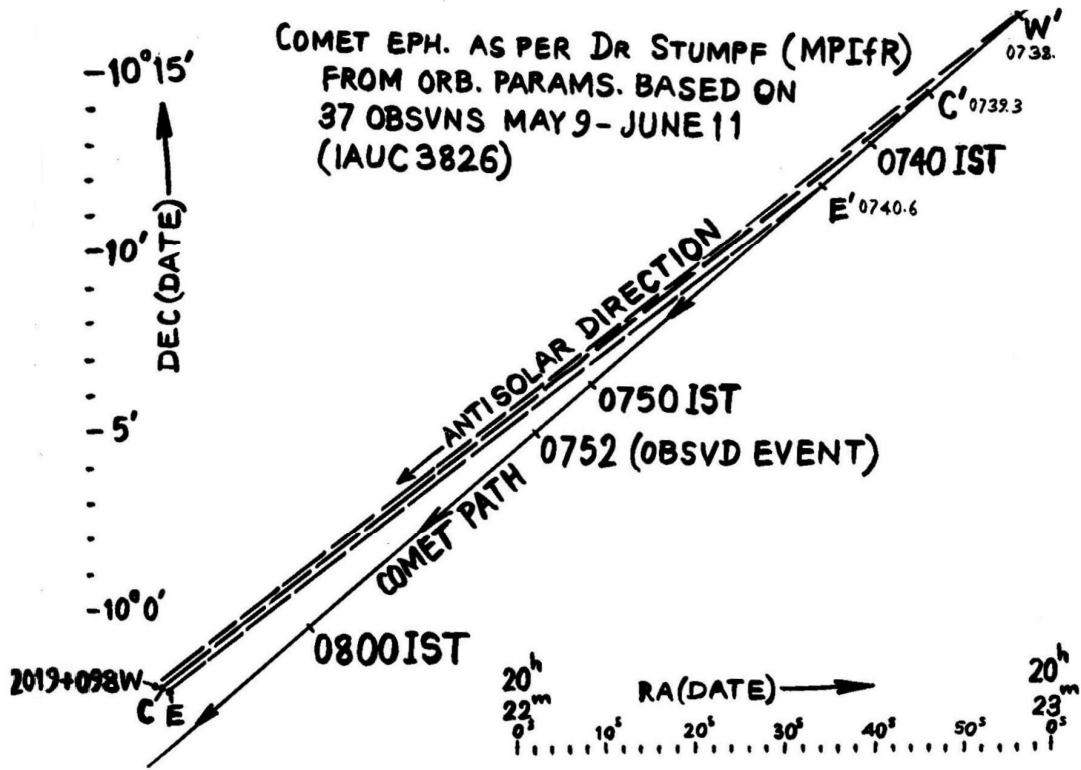

**Fig.1** Path of the comet in date coordinates as per ephemeris kindly provided by Dr Stumpf (MPIfR). W, C & E denote the western, central & eastern components of 3C411 and W', C' & E' the points at which straight lines from W, C & E along the expected tail direction cut the comet path.

-------------------------------------------

### 3. Results

We present in Fig.1 the path of the comet in date coordinates (right ascension and declination) as per ephemeris kindly provided by Dr. Stumpf of Max-Planck-Institute for Radio Astronomy (MPIfR), Bonn, who calculated it from the orbital parameters based on 37 observations during May 9 to June 11 (IAUC3826). W, C & E denote the western, central and eastern components of the radio source 2019+098 ≡ 3C411 (Appendix 2), and W', C' & E' the points at which straight lines from W, C & E along the expected tail direction cut the comet path.

--------------------------------------------------------

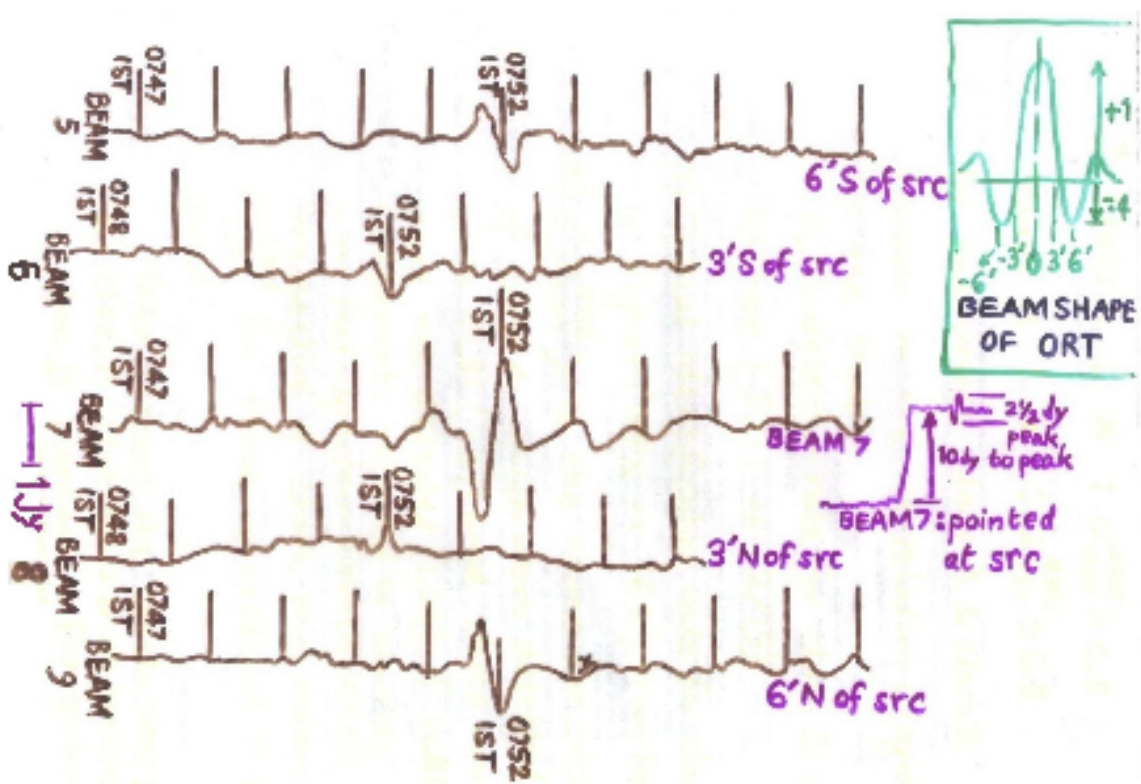

**Fig.2** Traces of correlated beams 5, 6, 7, 8 & 9 of the Ooty radio telescope (ORT) for a few minutes around the time of the observed event. Beam 7 is pointed at the source, beam 6 & 5 being 3 & 6 arcmin to the south and beams 8 & 9 being 3 & 6 arcmin to north. Beamshape of ORT is shown in the inset. The detected event at 07:52 IST clearly shows the beamshape (with slight correction needed for the pointing – see text).

--------------------------------------------------------

In Fig.2 we present traces of correlated beams 5, 6, 7, 8 & 9 of Ooty Radio Telescope (ORT) for a few minutes around the time of the observed event at 07:52 IST. Beam 7 is pointed at the source 2019+098, beams 6 & 5 being 3' & 6' to the south and beams 8 & 9 being 3' & 6' to north. From the north-south beamshape of the correlated ORT beams shown as an inset in the figure, beams 6 & 8 should have zero signal and beams 5 & 9 about -40% signal since the source is compact in the north-south (i.e., declination) (see Appendix 2). The traces in Fig.2 show the ORT beamshape very well. Beams 5 & 9 correctly showing about 40% signal of the opposite sign compared to the beam 7 signal. Beams 6 & 8 also show near zero signal, the small nonzero values may be used to correct the pointing in the north-south.

## 4. Solar wind velocity and comet velocity

### 4.1 Predicted occultation time

The comet consists of coma + ion tail. The tail can nominally be taken along solar wind, taken to be radially away from Sun. Corrections must then be applied for the difference in actual and nominal solar wind direction, and the comet velocity in relation to solar wind velocity. The first step in this procedure is to find the spherical angle NCP-comet-Sun.

From Sun's and comet's ephemeris for 12[th] June 1983, this angle is 52.9 deg. Nominally taking the ratio of solar wind to comet velocities to be 8 (say, 400 km/sec / 50 km/sec), the relevant angle for predicting occultation time becomes 52.5 deg, and varies between 49.5 to 52.7 deg as the velocity ratio goes from 0 (for stationary solar wind) to 15, for 08hr IST (+/- 20min). Predicted occultation time is sensitive to accuracy of determining relevant directions (i.e., Position Angles or Pas on sky), especially since directions source (i.e., 3C411) to Sun and comet to Sun are close to each other (52.9 and 49 deg respectively). Further, as shown in Fig.1, 3C411 has three distinct components, disposed more or less E-W (cf. Appendix 2). So although the nominal predicted occultation time is 07:24 IST, this prediction needs to be refined.

*4.2 A consistent model*
A consistent set of parameters need to be calculated to account for the nominal predicted occultation time 07:24 IST being different from 07:52 IST, at which the occultation was observed. … … …

*4.3 Parameters and uniqueness of the recommended model*

**5. Possible interpretations of the occultation profile**
Assuming that the geometry and kinematics account for the difference between predicted and observed occultation times, the occultation event is likely to be due to refraction of radiation from 3C411 through an ionized obstacle of optical depth << 1.

**Appendix 1**
Sources [S(408 MHz) ≥ 2 Jy] likely to be occulted by comet 1983e
(using ephemeris from IAUC3812)

| Source name | RA(1950) RA(Date) | Dec(1950) Dec(Date) | S(408) in Jy | Occultn date (June'83) | Predicted ET (=UT +50 sec) | Corresp. IST(=UT +5h30m) | Distance in dirn of tail (deg) |
|---|---|---|---|---|---|---|---|
| 1909-058 | 19   09 17.3 19   11 04.6 | -05 53 22 -05 50 03 | 2.93 | 12[th]/13th | 23h 29m | 04h 58m | 0.0 |
| 1909-071 | 19   09 44.5 | -07 11 15 | 2.26 | 13[th] | 01h 10m | 06h 40m | -2.0 |
| 1925-012 | 19   25 53.9 | -01 13 03 | 2.37 | 12[th]/13th | 19h 01m | 00h 31m | +1.6 |
| **2019+098 ≡ 3C411** | 20   19 44.5 20   21 20.9 | +09 51 32 +09 57 50 | 10.00 | 12[th] | 01h 54m | 07h 24m | 0.0 |
| 2023+119 | 20   23 00.2 | +11 56 29 | 2.42 | 12[th] | 00h 55m | 06h 25m | +2.2 |
| 2118+227 | 21   18 | +22 43 12 | 2.5 | 11[th] | 03h 36m | 09h 06m | +2.2 |

| | 25.9 | | | | | | |
|---|---|---|---|---|---|---|---|
| 2142+226 | 21 42 34.1 | +22 38 18 | 2.5 | 10th/11th | 22h 17m | 03h 47m | -4.0 |

Sources were selected from Molonglo, Parkes & 4C catalogues. For prediction of exact occultation time, the following points were considered.

(1) Exact ephemeris of comet was used, taking acceleration into account by using parabolic or cubic interpolation rather than just linear.

(2) Exact tail direction ≡ exact solar wind direction at the comet + effect of motion of comet ≈ line joining comet to Sun.

(3) Precession was done for the exact date.

(4) Parallax of comet (max 2.5 arcmin = radius of Earth / closest distance (in radians) = $6.4 \times 10^3 / 9 \times 10^6$ ) is important, although not aberration, which is the same for both comet and source.

**Appendix 2**
-----------------------------------------------------------------------

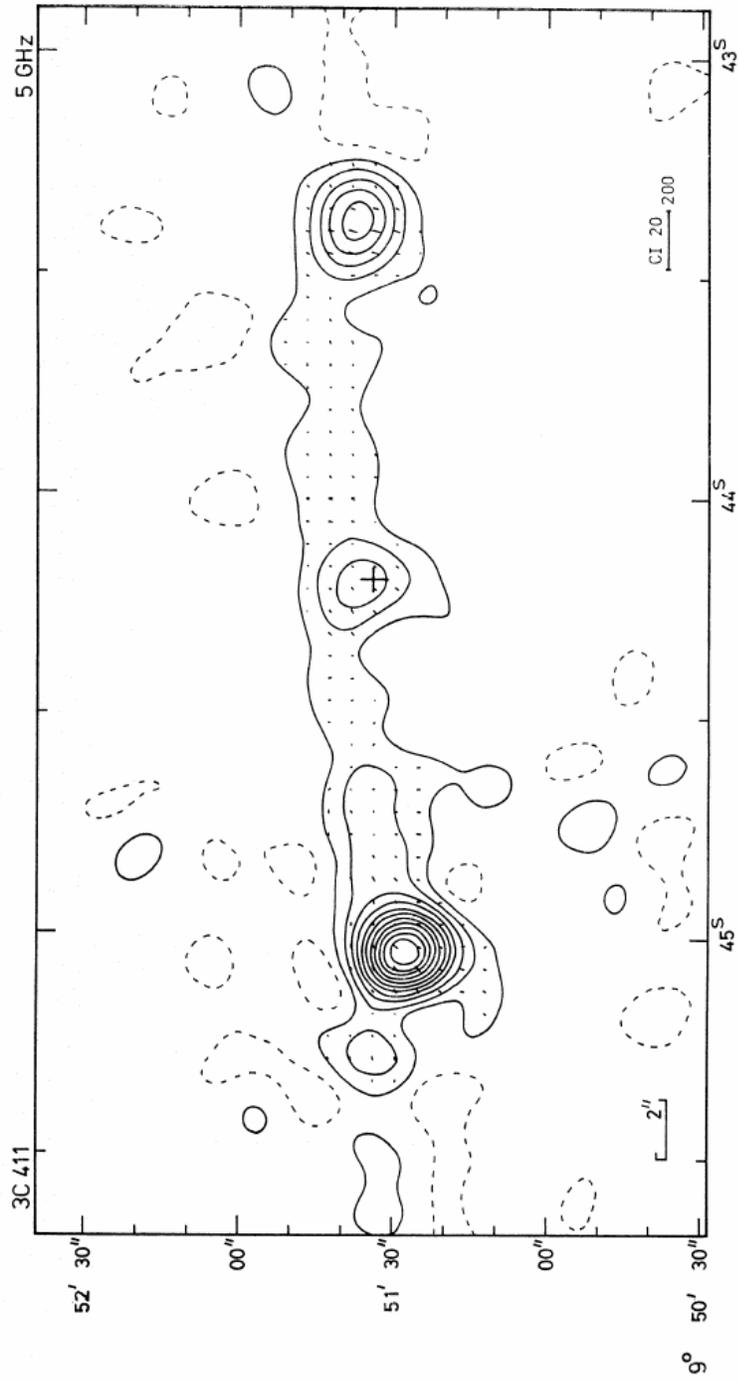

FIG. 2. 3C 411. This source is discussed in more detail by Spinrad et al. (1974), who also give the position of the 20ᵐ N galaxy marked.

**Appendix 3** <Info on comet 1983e = Sugano-Saigusa-Fujikawa)

**Acknowledgments**

We thank P K Manoharan for assistance in the observations.

**References** (= a bibliography of cometary occultation observations and related work)